# Flow-level Coordination of Connected and Autonomous Vehicles in Multilane Freeway Ramp Merging Areas

*Jie Zhu, Ivana Tasic, Xiaobo Qu*

*Department of Architecture and Civil Engineering, Chalmers University of Technology,*

*41296 Gothenburg, Sweden*

**Abstract**

On-ramp merging areas are deemed to be typical bottlenecks for freeway networks due to the intensive disturbances induced by the frequent merging, weaving, and lane-changing behaviors. The Connected and Autonomous Vehicles (CAVs), benefited from their capabilities of real-time communication and precise motion control, hold an opportunity to promote ramp merging operation through enhanced cooperation. The existing CAV cooperation strategies are mainly designed for single-lane freeways, although multilane configurations are more prevailing in the real-world. In this paper, we present a flow-level CAV coordination strategy to facilitate merging operation in multilane freeways. The coordination integrates lane-change rules between mainstream lanes, proactive creation of large merging gaps, and platooning of ramp vehicles for enhanced benefits in traffic flow stability and efficiency. The strategy is formulated under an optimization framework, where the optimal control plan is determined based on real-time traffic conditions. The impacts of tunable model parameters on the produced control plan are discussed in detail. The efficiency of the proposed multilane strategy is demonstrated in a micro-simulation environment. The results show that the coordination can substantially improve the overall ramp merging efficiency and prevent recurrent traffic congestions, especially under high traffic volume conditions.

*Keywords: coordinative ramp merging, connected and autonomous vehicles, multilane freeway, optimization, microscopic simulation*

# 1 Introduction

On-ramp merging areas are deemed to be typical bottlenecks for freeway networks, as the merging of ramp vehicles impose frequent disturbances on the traffic flow and cause various problems, such as traffic oscillations, increased energy use and pollutions, accidents, and recurrent traffic congestions (Cassidy et al., 1999, Mergia et al., 2013, Srivastava et al., 2013, Han et al., 2018, Wang et al., 2019). Many efforts are devoted to facilitating the merging operation at freeway on-ramps. Prior approaches mainly focus on the active traffic management strategies, such as ramp metering (Papageorgiou et al., 1991, Smaragdis et al., 2004, Gomes et al., 2006, Papamichail et al., 2010), variable speed limit (Carlson et al., 2011, Zhang et al., 2013, Chen et al., 2014), hard-shoulder running (Mirshahi et al., 2007), and the combinations of them (Hegyi et al., 2005, Papamichail et al., 2008, Carlson et al., 2010, Lu et al., 2011). However, the benefits of such systems are limited because they cannot address the randomness and heterogeneity in the microscopic dynamics of individual vehicles.

The emerging Connected and Autonomous Vehicles (CAVs) present an opportunity to regulate individual vehicles and achieve cooperative driving in various bottleneck areas (Zhou et al., 2017, Debada et al., 2018, Mirheli et al., 2019, Qu et al., 2020, Wu et al., 2020, Zhou et al., 2020, Cao et al., 2021, Wu et al., 2021, Zhu et al., 2021a). Compared to the Human-Driven Vehicles (HDVs), CAVs can obtain more accurate and comprehensive traffic information through the vehicle communication technologies, and they are capable of more precise and timely execution of the dynamic driving tasks. In the literature, a number of CAV-enabled cooperation strategies are proposed to facilitate the merging traffic at freeway on-ramps, as summarized in Scarinci et al. (2014) and Rios-Torres et al. (2017b). These strategies may differ in, for example, adopted assumptions, required vehicle capabilities (i.e., connected, autonomous, or connected and autonomous), penetration rate of "intelligence" (i.e., 100% CAVs or a mix of CAVs and HDVs), and hierarchy of control (i.e., centralized, decentralized, or combined), while sharing the common objective to improve merging operational performance. Most of these strategies focus on the lower-level decisions of individual vehicles (e.g., trajectory design of CAVs), whereas the upper-level considerations (e.g., traffic flow efficiency and stability) are only discussed to a very limited extent. Further, the existing studies are primarily designed for single-lane freeways where free lane-changes between mainstream lanes are neglected, whereas a comprehensive discussion on how to achieve merging cooperation in the more prevailing multilane freeways is missing.

In view of the limitations of existing CAV merging strategies, we propose in this paper an upper-level coordination strategy to promote the ramp merging operation in multilane freeways. The proposed strategy, as a significant extension to Zhu et al. (2021b), coordinates the ramp traffic with the mainline traffic in the outermost lane by proactively creating large gaps on the main road and leading ramp vehicles into the created gaps in the form of platoons. The free lane-changes between the mainstream lanes are also considered in the coordination design, and the mainline-ramp coordination is combined with the one-sided lane-change prohibition rule to protect the created outer lane gaps from being occupied by the inner lane vehicles. The coordination is formulated as an optimization problem based on microscopic and macroscopic traffic flow models. The formulation clearly considers traffic stability and efficiency gains at the continuous traffic flow level. The benefits of the proposed coordination are demonstrated through a case study conducted on a microscopic simulation platform.

The remaining of this paper is structured as follows. Section 2 reviews the state-of-the-arts of CAV ramp merging strategies and summarizes the contributions of this work. Section 3 formulates the multilane coordination strategy and provides extended discussions on relevant issues. Section 4 introduces the case study and discusses the efficiency of CoMC under various traffic conditions. The conclusion is drawn in Section 5.

## 2 Literature Review

CAV ramp merging strategies, deemed to be a promising approach to facilitate traffic operation in the freeway on-ramp bottlenecks, have received great attention in the recent research efforts. The existing efforts can be mainly divided into two categories: trajectory planning methods and aggregated control methods.

A majority of the existing studies falls into the first category. These studies usually use an optimization framework to formulate the lower-level motion plan of multiple vehicles, subject to vehicle dynamics, safety requirements, and technical constraints, but major differences may lie in the control direction (e.g., longitudinal, lateral, or both), the objective (e.g., efficiency, safety, energy use, passenger comfort, or combined cost), and the control variables (e.g., acceleration, jerk, and/or lane-change decisions). For example, Cao et al. (2015) describe the states and actions of a ramp merging vehicle and its mainline competitor in a two-dimensional coordinate system and design their optimal paths, defined by the longitudinal acceleration and the optimal merging point, by minimizing a penalty combined of acceleration, speed deviation, and inter-vehicle distance. Similarly, Zhou et al. (2019a) and Zhou et al. (2019b) solve the merging trajectories of a ramp merging vehicle and a mainline facilitating vehicle at a minimal acceleration cost, while restraining the negative safety impacts of the merging maneuver. Karimi et al. (2020) look at the mixed CAV-HDV traffic flow and control the cooperative motions of CAVs for different types of merging triplets. The above studies focus on the interaction between a ramp merging vehicle and its direct neighbors on the main road, whereas some other studies assume the presence of an upper-level merging sequence and jointly plan the trajectories of a series of vehicles within the merging control zone. For example, Ntousakis et al. (2016), Rios-Torres et al. (2017a) and Sonbolestan et al. (2021) minimize the overall acceleration/jerk efforts in favor of energy use and passenger comfort. The models are analytically solved using Hamiltonian analysis. Letter et al. (2017) and Xie et al. (2017)

formulate models targeting at maximum collective speed, while taking into account the safety distance between vehicles. The strategy of Letter et al. (2017) is later extended to a two-lane freeway configuration, where a centralized lane-changing controller is integrated to reallocate a number of mainstream vehicles from the outer lane to the inner lane (Hu et al., 2019). Further, Omidvar et al. (2020) apply Letter et al. (2017) in a mixed CAV-HDV condition, where a model predictive control framework is used to address the deviations in HDV behaviors. Some recent studies integrate the choice of merging sequence into the motion planning problem. For example, Ding et al. (2020) use a series of rules to adjust the merging order of vehicles and then plan the motion of each vehicle accordingly. Xu et al. (2021) formulate the merging sequence choice as an optimization problem combining the mainline travel time and the merging throughput and solve the problem with generic algorithms. Alternatively, Jing et al. (2019), Chen et al. (2020) and Sun et al. (2020) integrate the gap choice and path planning of merging vehicles into a single optimization model. The model compares the path costs to lead a ramp vehicle into different gaps, so as to find the optimal gap and the corresponding path with the lowest cost. In addition to the optimization method, alternative approaches to determine vehicle trajectories at ramp merging are also proposed. For example, Marinescu et al. (2012) and Wang et al. (2013) utilize the idea of virtual vehicle/slot to reserve mainline space for the merging vehicles. Fukuyama (2020) develop a dynamic game-based framework where each vehicle chooses the most beneficial action based on the estimated action of a competing vehicle. Karbalaieali et al. (2020) considers a two-lane freeway and choose from various alternative actions (i.e., speeding up, slowing down, or changing lanes) the one that minimizes the total travel time of a ramp vehicle and its mainline competitors. In summary, the above studies are dedicated to facilitating the ramp merging operation through joint trajectory planning of relevant vehicles. Though presenting encouraging results, these methods mainly focus on the microscopic interactions between individual vehicles, whereas their performance in a continuous traffic flow is not ensured.

The second category, aggregated control methods, is less covered in the literature. It refers to the ramp merging strategies that control the traffic flow operation, instead of the step-by-step action of individual vehicles, at a more aggregated level. An example is Scarinci et al. (2017), where the mainline traffic is periodically compacted to create large merging gaps, and the ramp traffic is released into the gaps via a ramp metering signal. This strategy can ensure the fluent operation of mainline traffic, but arising a fairness concern between traffic streams, as the release of ramp traffic is fully dependent on the main road condition under such a system. In Chen et al. (2021), a similar idea of periodic gap creation is adopted and combined with a batch merging strategy to close the extra gaps induced by lane-changes. The benefits of the proposed system are demonstrated in theory, but no numerical or simulation experiment is carried out. Recently, Zhu et al. (2021b) develop a coordination strategy that creates on-demand gaps on the main road upon formations of proper ramp merging platoons. The coordination determines adaptive control plan according to the real-time traffic state. Case study results show the strategy's ability to improve traffic efficiency and prevent congestions at ramp merging. It is noticed in the review that the existing flow-level coordination strategies only consider the single-lane freeway configuration and ignore the free lane-changes between mainstream lanes, so a thorough discussion about how to implement merging coordination in a multilane freeway configuration is desired.

In summary, in the current literature, only very few CAV merging strategies are designed for the multilane freeway environment, such as Hu et al. (2019) and Karbalaieali et al. (2020), and the existing multilane strategies mainly focus on the lower-level decisions of individual vehicles, whereas the options to improve traffic flow performance are not sufficiently discussed. Therefore, we are motivated to propose in this paper a flow-level strategy, enabled by the CAV capabilities, to coordinate the two streams of traffic (instead of individual vehicles) in the multilane freeway ramp merging areas.

# 3 Coordinative Merging Control (CoMC) in Multilane Freeway

3.1 Formulation of multilane CoMC

The Coordinative Merging Control (CoMC) strategy for multilane freeways is developed based on the coordination framework proposed in Zhu et al. (2021b). The underlying idea is to coordinate the outer-lane mainline traffic and the ramp traffic through proactive gap creation on the main road and platoon formation on the ramp. Specifically, as shown in Fig. 1, the on-ramp vehicles should stop at a pre-determined Waiting Position (WP) on the ramp and register themselves with the control center upon arrival. The control center counts the number of vehicles waiting on the ramp and initiates a coordinative merging request when a certain number of ramp vehicles has accumulated. With the merging request, the control center will first appoint a vehicle in the mainstream outer lane as the facilitating vehicle and require it to slow down at the Speed-Change (SC) position, so that a gap is proactively created between the facilitating vehicle and its leader on the main road. Then, the control center releases the vehicles waiting on the ramp as a platoon by specifying their moving trajectories towards the Merging Point (MP). In order to smoothly guide the merging platoon into the created gap, the centralized control should make a joint decision on the control variables, including the SC position and cooperative speed of the facilitating vehicle and the size and moving trajectory of the merging platoon, so as to satisfy the following three requirements at merging: (1) the created mainline gap should be large enough for the merging platoon (the requirement of size); (2) the platoon should reach the same speed as the mainline facilitating vehicle at the MP (the requirement of merging speed); and (3) the gap should be just available at the MP when the platoon arrives there (the requirement of arrival time). This entire course of creating a gap and guiding a merging platoon into the gap is defined as a coordinative merging cycle, and the CoMC system functions by recurrently implementing the merging cycles.

Note that, the essence of such a coordination system is to compact the mainline traffic by reducing the traffic speed, thereby collecting enough space for the merging of ramp vehicles. We assume that the mainline traffic is originally in a stable equilibrium state (state O) before the coordination is initiated. When the facilitating vehicle decelerates in response to a coordination request, the mainline vehicles following the facilitating vehicle (i.e., cooperative vehicles in Fig. 1) should also slow down and accept a shorter car-following distance corresponding to the reduced speed. This transfers the mainline traffic behind the facilitating vehicle into a cooperative state (state C), characterized by a higher density and an increased flow rate in comparison to the original state. The transition of mainline traffic state provides space for the merging traffic; however, it also generates a shockwave that negatively affects the mainline traffic. If the coordination is initiated too frequently, the shockwaves may accumulate and eventually trigger traffic breakdowns in the merging area. To this end, it should be ensured in the coordination that the merging traffic is facilitated without breaking the mainline stability.

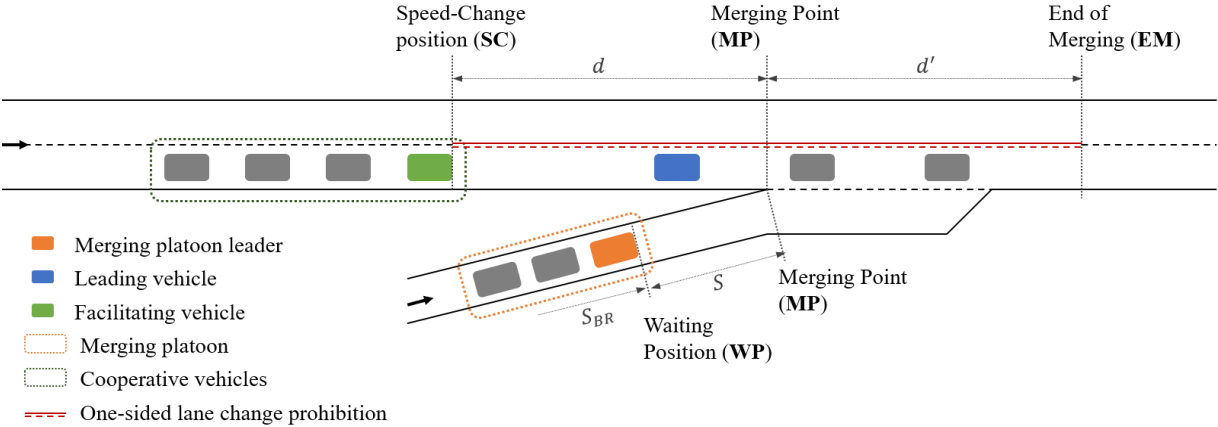

Fig. 1 Conceptual illustration of the CoMC system

Further, in a multilane freeway, vehicles in the inner lanes may change into the gaps created in the outer lane, if no additional control is applied, leading to a failure of the mainline-ramp coordination. Therefore, we recommend combining CoMC with the one-sided lane-change prohibition rule, which allows vehicles in the outermost lane to change into the inner lanes (the facilitating vehicle should not change lanes), while prohibiting vehicles in the inner lanes from entering the outermost lane. The lane-change prohibition should cover the entire control segment (i.e., from the speed-change position to the end-of-merging position as shown in Fig. 1) and be effective during the whole coordination period. This measure can prevent the created gaps from being occupied by the inner lane vehicles and at the meantime speed up the dissipation of shockwaves by allowing the outer lane vehicles to change lanes.

The following assumptions are applied throughout this work:

- 100% penetration rate of CAVs in the freeway network.
- All vehicles are highly automated corresponding to SAE (2016) level 4.
- The control center and relative vehicles are capable of real-time communication through vehicle to infrastructure (V2I) technologies.

Note that, in each coordinative cycle, CoMC only controls the facilitating vehicle and the merging platoon leader, whereas the other vehicles can obey the regular car-following rules. The main purpose of requiring a 100% CAV penetration rate is to ensure that any vehicle in the network can take the role of facilitating vehicle (for mainline vehicles) or platoon leader (for ramp vehicles). In addition, as CoMC adopts a centralized control framework, vehicle to vehicle (V2V) communication is not required, although it may benefit the system performance in terms of more efficient information transmission.

The CoMC strategy is analytically formulated as a constrained optimization problem, incorporating macroscopic and microscopic traffic flow models, as defined through (1)-(15) with the notation in Table 1. The model minimizes the total delay to all vehicles passing through the ramp merging area, including the mainline cooperative vehicles and the ramp vehicles. Three decision variables are controlled in the model, i.e., the merging platoon size $n$, the mainline cooperative distance $d$, and the cooperative speed $v_c$. The requirements on coordination, traffic stability, and vehicle dynamics are explicitly considered through the model constraints, where (7) requires the created mainline gap to be no smaller than the space needed for the merging platoon, (8) and (9) stipulate that the mainline traffic should not break due to the cooperative behaviors, (10) ensures a feasible acceleration rate of the merging platoon, and (14)-(15) describe the nature of the decision variables. In addition, this model expresses $q_C$, $k_C$, and $h_C$ as functions of $v_C$, because the relationships between these variables are essentially defined by the fundamental diagram of traffic flow, which can be fitted to field data or derived from a recognized car-following model (Jin, 2016). Theoretically, CoMC is compatible with any form of fundamental diagram. Without loss of generality, we use the fundamental diagram derived from the Wiedemann 99 car-following model (Wiedemann, 1991) in this paper, as defined in (11)-(13). We refer to Zhu et al. (2021b) for detailed derivations and explanations on the model formulation and the solution methods.

$$min\ D = \left( w_m \cdot \sum_{i=1}^{m} D_{main}^i + w_r \cdot \sum_{j=1}^{n} D_{ramp}^j \right) \times r \quad (1)$$

with

$$\sum_{i=1}^{m} D_{main}^i = \frac{m \cdot (v_O - v_C)}{v_C} \times \left[ \frac{d + d'}{v_O} - \frac{(m-1)\omega h_O}{2(v_O - \omega)} \right] \quad (2)$$

$$\sum_{j=1}^{n} D_{ramp}^j = n \times \left( \frac{v_r}{2b} + \frac{d + d'}{v_C} - nh_C - \frac{d - nh_C v_C}{2v_r} - \frac{d'}{v_O} + \frac{n-1}{2\lambda} \right) \quad (3)$$

$$r = \frac{3600\lambda}{n} \quad (4)$$

$$m = \left\lceil \frac{d + d'}{h_O} \times \left( \frac{1}{\omega} - \frac{1}{v_O} \right) \right\rceil \quad (5)$$

$$\omega = \frac{q_C - q_O}{k_C - k_O} \quad (6)$$

subject to

$$h_O + \frac{d}{v_C} - \frac{d}{v_O} \geq (n + 1) \cdot h_C \quad (7)$$

$$\frac{n}{\lambda} \geq \frac{d + d'}{\omega} \tag{8}$$

$$v_{crit} \leq v_C \leq v_O \tag{9}$$

$$a \leq a_{max} \tag{10}$$

$$h_C = \frac{CC0 + L + CC1 * v_C}{v_C} \tag{11}$$

$$q_C = 1/h_C \tag{12}$$

$$k_C = q_C/v_C \tag{13}$$

$$n \leq n_{max}, n \in \mathbb{N}^+ \tag{14}$$

$$d > 0 \tag{15}$$

Table 1 Notation

| Variable | Description | Role |
| --- | --- | --- |
| $n$ | Merging platoon size | decision variable |
| $d$ | Cooperative distance (i.e., distance between SC and MP) | decision variable |
| $v_C$ | Mainline cooperative speed | decision variable |
| $w_m$ | Weight of mainline traffic | input parameter |
| $w_r$ | Weight of ramp traffic | input parameter |
| $D_{main}^i$ | Delay to the $i^{th}$ mainline cooperative vehicle | function of $v_C$ and $d$ |
| $D_{ramp}^j$ | Delay to the $j^{th}$ ramp vehicle in the merging platoon | function of $n$, $v_C$, and $d$ |
| $r$ | frequency of the merging cycles | function of $n$ |
| $q_O$ | Mainline flow rate in the original state | input parameter |
| $k_O$ | Mainline density in the original state | input parameter |
| $v_O$ | Mainline vehicle speed in the original state | input parameter |
| $h_O$ | Mainline headway in the original state | input parameter |
| $q_C$ | Mainline flow rate in the cooperative state | function of $v_C$ |
| $k_C$ | Mainline density in the cooperative state | function of $v_C$ |
| $h_C$ | Mainline headway in the cooperative state | function of $v_C$ |
| $m$ | Number of mainline cooperative vehicles | function of $v_C$ and $d$ |
| $\omega$ | Shockwave speed | function of $v_C$ |
| $d'$ | Distance between MP and EM | input parameter |
| $v_r$ | Arrival speed of ramp vehicles | input parameter |
| $v_{crit}$ | Critical speed of mainline traffic | input parameter |
| $\lambda$ | Arrival rate of ramp vehicles | input parameter |
| $b$ | Braking rate of ramp vehicles when approaching WP | input parameter |
| $a_{max}$ | Maximum allowable acceleration of ramp vehicles | input parameter |
| $n_{max}$ | Maximum length of a merging platoon | input parameter |

Note that, in a multilane configuration, the ramp traffic is coordinated with the mainline traffic in the outer lane. Thus, the traffic volume in the outer lane should be estimated and used as the original flow rate of coordination ($q_O$). Under the CoMC control, when the facilitating vehicle decelerates, the vehicles following it in the outer lane may tend to change into the inner lanes to maintain a higher speed. We assume a two-lane freeway where the traffic volume in the upstream road segment is evenly distributed between lanes (see Fig. 2), and the number of vehicles changing into the inner lane depends on the ability of the inner lane to accommodate extra vehicles. The remaining outer lane volume for coordination is estimated as

$$q_O = q_m - \rho \cdot (C - q_m) \tag{16}$$

where $q_m$ is the upstream mainline volume per lane, $C$ is the theoretical capacity of the inner lane which is defined by the fundamental diagram of traffic flow, and $\rho \epsilon [0,1]$ captures the number of lane-changing vehicles as a fraction of the reserved inner lane capacity (i.e., $C - q_m$). The impacts of $\rho$ is discussed in detail in Section 3.3.

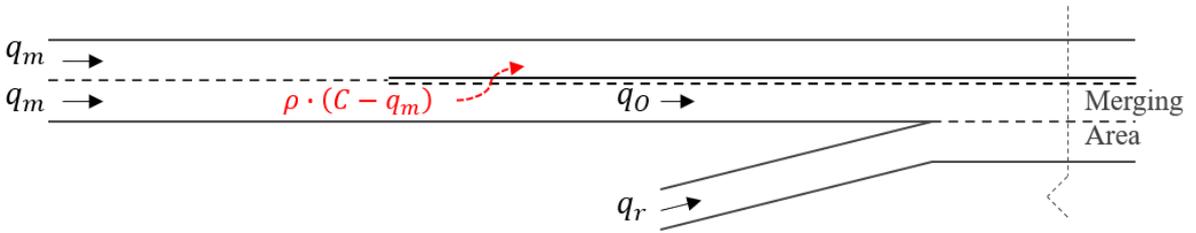

Fig. 2 Effective outer lane flow rate

3.2 Impacts of weight choice

The weight parameters $w_m$ and $w_r$ adjust the priority of the mainline and ramp traffic in the coordination design and have an impact on the optimal CoMC control plan. We conduct a sensitivity analysis where the values of $w_m$ and $w_r$ vary from 0 to 1 (with $w_m + w_r = 1$) for the parameters in Table 2. The results for two demand scenarios (mainline volume 2000 and 2200 veh/h/ln with ramp volume 500 veh/h) are presented in Table 3.

Table 2 Input parameter

| Parameter | Value | Unit | Source |
|---|---|---|---|
| $v_O$ | 120 | km/h | TRB (2016) |
| $v_r$ | 60 | km/h | TRB (2016) |
| $d'$ | 457.2 | m | TRB (2016) |
| $v_{crit}$ | 75 | km/h | Geistefeldt et al. (2017) |
| $b$ | 2.75 | m/s² | AASHTO (2018) |
| $a_{max}$ | 2.75 | m/s² | AASHTO (2018) |
| $n_{max}$ | 20 | veh | - |
| $\rho$ | 0.5 | - | - |
| $CC0$ | 1.5 | m | PTV (2018) |
| $CC1$ | 0.9 | s | PTV (2018) |
| $L$ | 4.37 | m | PTV (2018) |

Table 3 Sensitivity analysis on weight parameters

(a) CoMC control plan for mainline volume 2000 veh/h/ln and ramp volume 500 veh/h

| $w_m$ | 0.0 | 0.1 | 0.2 | 0.3 | 0.4 | 0.5 | 0.6 | 0.7 | 0.8 | 0.9 | 1.0 |
|---|---|---|---|---|---|---|---|---|---|---|---|
| $w_r$ | 1.0 | 0.9 | 0.8 | 0.7 | 0.6 | 0.5 | 0.4 | 0.3 | 0.2 | 0.1 | 0.0 |
| $v_C$ (km/h) | 85.4 | 85.4 | 85.4 | 85.4 | 85.4 | 85.4 | 85.4 | 85.4 | 85.4 | 88.0 | 89.6 |
| $d$ (m) | 1044 | 1044 | 1044 | 1044 | 1044 | 1044 | 1044 | 1044 | 1044 | 1260 | 1458 |
| $n$ (veh) | 11 | 11 | 11 | 11 | 11 | 11 | 11 | 11 | 11 | 12 | 13 |

(b) CoMC control plan for mainline volume 2200 veh/h/ln and ramp volume 500 veh/h

| $w_m$ | 0.0 | 0.1 | 0.2 | 0.3 | 0.4 | 0.5 | 0.6 | 0.7 | 0.8 | 0.9 | 1.0 |
|---|---|---|---|---|---|---|---|---|---|---|---|
| $w_r$ | 1.0 | 0.9 | 0.8 | 0.7 | 0.6 | 0.5 | 0.4 | 0.3 | 0.2 | 0.1 | 0.0 |
| $v_C$ (km/h) | 81.1 | 81.1 | 81.1 | 81.1 | 81.1 | 81.1 | 81.1 | 81.1 | 84.8 | 84.8 | 86.6 |
| $d$ (m) | 1136 | 1136 | 1136 | 1136 | 1136 | 1136 | 1136 | 1136 | 1391 | 1391 | 1593 |
| $n$ (veh) | 14 | 14 | 14 | 14 | 14 | 14 | 14 | 14 | 15 | 15 | 16 |

As Table 3 shows, when the mainline traffic is prioritized (i.e., with larger $w_m$), CoMC tends to form larger merging platoons to reduce the frequency of coordination, and the mainline cooperative speed is higher in comparison with lower mainline priority. In addition, it is noted for both demand levels that the control plan remains the same when $w_m$ changes between 0 and 0.7, implying that CoMC tends to prioritize the ramp efficiency when the mainline and ramp traffic have similar weights. For example, see Table 3, the case $w_m = w_r = 0.5$ has the same control plan as the case $w_m = 0$ for both demand levels. This is because under such a coordination system where the ramp vehicles are forced to stop and give way to the mainline vehicles, the ramp traffic inherently experiences larger delay in comparison to the mainline traffic, so the control tends to minimize the dominating ramp delay for an improvement in the overall efficiency. In the following analysis, we use $w_m = w_r$ to treat each vehicle (regardless of whether it is from mainline or on-ramp) evenly.

3.3 Impacts of cooperative lane changes

The parameter $\rho$ describes the proportion of the reserved inner lane capacity that is utilized by the vehicles changing from the outer lane. Its value can affect the effective outer lane flow rate in the original state ($q_O$), and thus affecting the control decisions of CoMC. Fig. 3 shows the maximum on-ramp flow that can be accommodated by CoMC with respect to the mainline volume for the parameters in Table 2 and $w_m = w_r$. With a larger value of $\rho$ (i.e., more outer lane vehicles changing to the inner lane at the coordination), more space can be collected in the outer lane to facilitate the merging of ramp vehicles, so the maximum accommodated on-ramp flow increases. For example, the maximum ramp flow increases from 551 veh/h to 690 veh/h (approximately 25.2 %) when $\rho$ increases from 0.2 to 0.8 at a mainline volume of 2000 veh/h/ln.

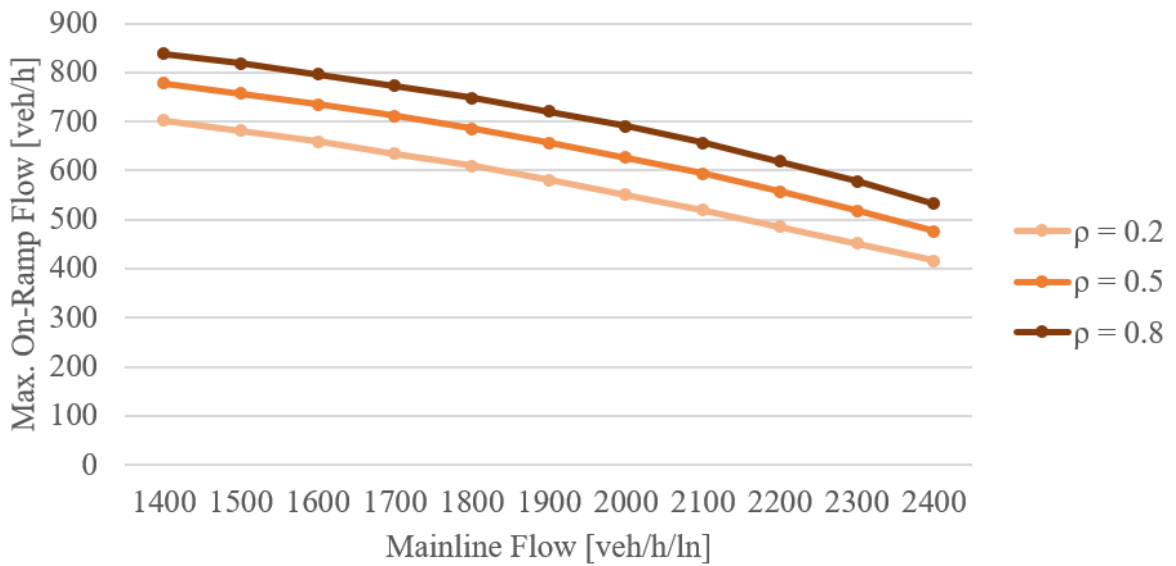

Fig. 3 Maximum on-ramp flow with respect to mainline flow

Table 4 presents the CoMC control plan of an example scenario (mainline volume 2000 veh/h/ln and ramp volume 500 veh/h) under different values of $\rho$. It shows that the required merging platoon size ($n$) decreases as $\rho$ increases. This is because when more vehicles change into the inner lane, the shockwave in the outer lane can dissipate faster, and the coordination can be implemented more frequently (i.e., at the formation of a smaller merging platoon with fewer vehicles). The frequent coordination can improve the efficiency of ramp vehicles by reducing their waiting time on the ramp. In addition, since a smaller merging platoon requires less space on the main road, the facilitating vehicle can decelerate at a later SC position (i.e., smaller value of $d$) and maintain a higher cooperative speed (i.e., larger value of $v_C$) with the larger $\rho$ value.

Table 4 CoMC control plan for different $\rho$ values (mainline volume 2000 veh/h/ln, ramp volume 500 veh/h, $w_m = w_r$)

| $\rho$ | 0.0 | 0.1 | 0.2 | 0.3 | 0.4 | 0.5 | 0.6 | 0.7 | 0.8 | 0.9 | 1.0 |
|---|---|---|---|---|---|---|---|---|---|---|---|
| $v_C$ (km/h) | No solution | 80.1 | 83.5 | 83.3 | 85.0 | 85.4 | 84.9 | 82.9 | 86.1 | 82.8 | 85.5 |
| $d$ (m) | | 1331 | 1324 | 1142 | 1120 | 1044 | 930 | 781 | 878 | 692 | 765 |
| $n$ (veh) | | 17 | 15 | 13 | 12 | 11 | 10 | 9 | 9 | 8 | 8 |

In summary, a larger value of $\rho$ implies more changes from the outer lane to the inner lane, which can promote the merging operation because it essentially makes fuller use of the inner lane space for the merging of ramp vehicles. A low value of $\rho$ may lead to a waste of the inner lane capacity and even result in failed coordination (e.g., in Table 4, the model has no solution for $\rho = 0$). On the other hand, an excessively high value of $\rho$, implying frequent changes into the inner lane, may overload the inner lane and break the stability of the upstream traffic. Thus, it is important to determine a reasonable $\rho$ value for the balance between the merging efficiency and the mainline stability. Note that, with the presence of a centralized control system, it is possible to centrally determine the optimal value of $\rho$ and plan the cooperative lane changes based on the traffic conditions. Such a system is part of our on-going research.

**4 Case Study**

In this section, we conduct an illustrative case study through microscopic simulation to demonstrate the efficiency of CoMC in a multilane freeway configuration. Section 4.1 introduces the simulation design, and the results and discussions are provided in Section 4.2.

## 4.1 Simulation experiment

The experiments are conducted on a VISSIM-based microscopic simulation platform, where the internal VISSIM model is used to create the road network, generate traffic demands, control normal following/lane-changing decisions of vehicles, and record raw data for the performance analysis (PTV, 2018). It is shown that the default VISSIM following/lane-changing models are capable of reproducing realistic traffic dynamics at on-ramp merging (Marinescu et al., 2012, Scarinci et al., 2017, Xie et al., 2017, Hu et al., 2019). The centralized control of CoMC is compiled in Python and called by VISSIM via the COM interface. Specifically, VISSIM passes real-time traffic information (e.g., path, position, and speed of vehicles) to Python at each time step, and Python calculates the decisions of CoMC accordingly and return them to VISSIM. When the coordination is in progress, the dynamics of the facilitating vehicle and the merging platoon leader (i.e., the required speed adaption) is controlled by external driving models coded in C++. The activation and deactivation of external driving models are also part of the decisions made in the Python script.

Fig. 4 shows the simulation network of a merging area with two lanes on the freeway and one-lane on the ramp. The freeway consists of a 2000-meter-long upstream segment, a 240-meter-long merging area, and a 500-meter-long downstream segment. A 700-meter-long on-ramp is connected to the merging area via a parallel acceleration lane. Only one-direction traffic is modelled, as the opposing traffic would simply operate in a symmetrical way. One-sided lane-change prohibition is applied in and near the merging area as introduced in Section 3.1.

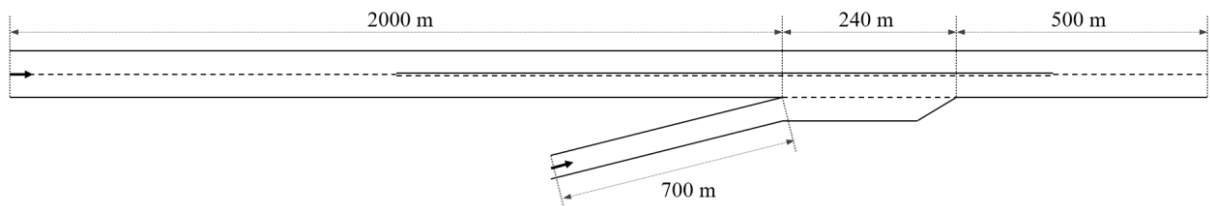

Fig. 4 Graphical representation of the simulation network

The performance of CoMC is evaluated in six demand scenarios with relatively high traffic demand, combing two levels of mainstream volume (2000 and 2200 veh/h/ln) and three levels of on-ramp volume (300, 400, and 500 veh/h). The scenarios and the corresponding CoMC control plans, solved for parameters in Table 2 and $w_m = w_r$, are summarized in Table 5. Note that, we only consider high traffic volume scenarios, because CoMC is dedicated to stabilizing traffic and promoting merging in the critical near-capacity situations. Under low demand situations where the traffic is well self-sustained, a centralized control like CoMC may not be needed. Relevant issues on this point is discussed in detail in Zhu et al. (2021b). For each demand scenario, we conduct 10 simulation runs (each lasting 7200 simulation seconds) with the centralized control of CoMC (i.e., the CoMC case) and 10 without CoMC (i.e., the base case). The simulation runs of the same case use different random seeds. Based on the results aggregated over 10 runs, the efficiency of the CoMC and base cases are compared to each other in terms of travel time, delay, and vehicle speed.

Table 5 Simulation scenarios and CoMC control plan

| Scenario | | 1A | 1B | 1C | 2A | 2B | 2C | Unit |
|---|---|---|---|---|---|---|---|---|
| Volume | $q_m$ | 2000 | 2000 | 2000 | 2200 | 2200 | 2200 | veh/h/ln |
|  | $q_r$ | 300 | 400 | 500 | 300 | 400 | 500 | veh/h |
| CoMC control | $v_C$ | 98.5 | 92.9 | 85.4 | 100.0 | 90.1 | 81.1 | km/h |
|  | $d$ | 687 | 909 | 1044 | 934 | 917 | 1137 | m |
|  | $n$ | 4 | 7 | 11 | 5 | 8 | 14 | veh |

To reproduce the fluctuating nature of traffic flow, vehicles are generated at randomized intervals in the simulation. Therefore, the original distance between a facilitating vehicle and its leader on the main road is different across merging cycles. When a large original gap already exists before the coordination, the facilitating vehicle can decelerate at a later position and still expand the gap to the required size. The deferred deceleration can reduce the delay to the mainline vehicles without affecting the effectiveness of coordination. In view of this, we introduce a microscopic mechanism to fine-tune the actual speed-change position in each merging cycle according to the initial position of the facilitating vehicle, as in (17).

$$d^* = d - \frac{(P_f - d)v_C}{v_f - v_C} \tag{17}$$

Here, $d^*$ is the actual speed-change position defined by its distance to the merging point. $P_f$ and $v_f$ are the position and speed of the facilitating vehicle when it accepts the coordination request. During the simulation, CoMC always appoint the first mainline vehicle behind the SC position as the facilitating vehicle, so $P_f \geq d$ always hold. At the micro-level, $d$ and $v_C$ are determined by the macroscopic CoMC control plan and remain unchanged across merging cycles, whereas $P_f$ and $v_f$ are updated in each merging cycle to determine $d^*$. This adaption can enhance the adaptivity of CoMC by taking into account traffic variations at the micro-level.

4.2 Result and discussion

Table 6 and Fig. 5 present the travel time and delay results. Travel time is measured as the total time that a vehicle takes to pass across the road network, except for the first 100 meters of the upstream links and the last 100 meters of the downstream link. Delay is the difference between the measured travel time and the theoretical ideal travel time that corresponds to the length and design speed of links in the vehicle path. The travel time and delay are measured separately for the mainline and ramp vehicles, so as to reveal the different effects of CoMC on vehicles in different paths. The results over all vehicles are also reported to indicate the overall traffic efficiency. According to the results, the effects of CoMC is less remarkable in the scenarios with relatively low volumes, such as 1A and 1B, because the traffic only experience minor delays (overall less than 10 seconds) even without external control. For scenarios 1A to 2B, CoMC mainly improves the efficiency of ramp traffic, because in the uncontrolled base cases, the ramp vehicles can hardly find acceptable merging gaps due to the high mainline volume, whereas CoMC solves this problem by creating readily available gaps at the merging point. For the most critical scenario 2C, CoMC is shown to improve both the mainline and ramp efficiency substantially with an 86.4% reduction in the overall delay in comparison to the base case.

Table 6 Travel time and delay results

| Scenario | | Mainline travel time [s] | Ramp travel time [s] | Overall travel time [s] | Mainline delay [s] | Ramp delay [s] | Overall delay [s] |
|---|---|---|---|---|---|---|---|
| 1A (2000,300) | base | 77.89 | 89.35 | 78.68 | 1.69 | 34.15 | 3.93 |
| | CoMC | 77.43 | 76.46 | 77.36 | 1.23 | 21.26 | 2.63 |
| 1B (2000,400) | base | 79.62 | 97.16 | 81.22 | 3.42 | 41.96 | 6.93 |
| | CoMC | 77.72 | 86.69 | 78.53 | 1.52 | 31.49 | 4.24 |
| 1C (2000,500) | base | 83.10 | 106.45 | 85.70 | 6.90 | 51.25 | 11.83 |
| | CoMC | 78.41 | 97.05 | 80.48 | 2.21 | 41.85 | 6.61 |
| 2A (2200,300) | base | 81.18 | 115.29 | 83.35 | 4.98 | 60.09 | 8.49 |
| | CoMC | 78.41 | 84.22 | 78.78 | 2.21 | 29.02 | 3.92 |
| 2B (2200,400) | base | 88.82 | 124.13 | 91.67 | 12.62 | 68.93 | 17.17 |
| | CoMC | 79.66 | 91.93 | 80.68 | 3.46 | 36.73 | 6.23 |
| 2C (2200,500) | base | 155.52 | 129.74 | 152.83 | 79.32 | 74.54 | 78.82 |
| | CoMC | 81.90 | 110.25 | 84.78 | 5.70 | 55.05 | 10.72 |

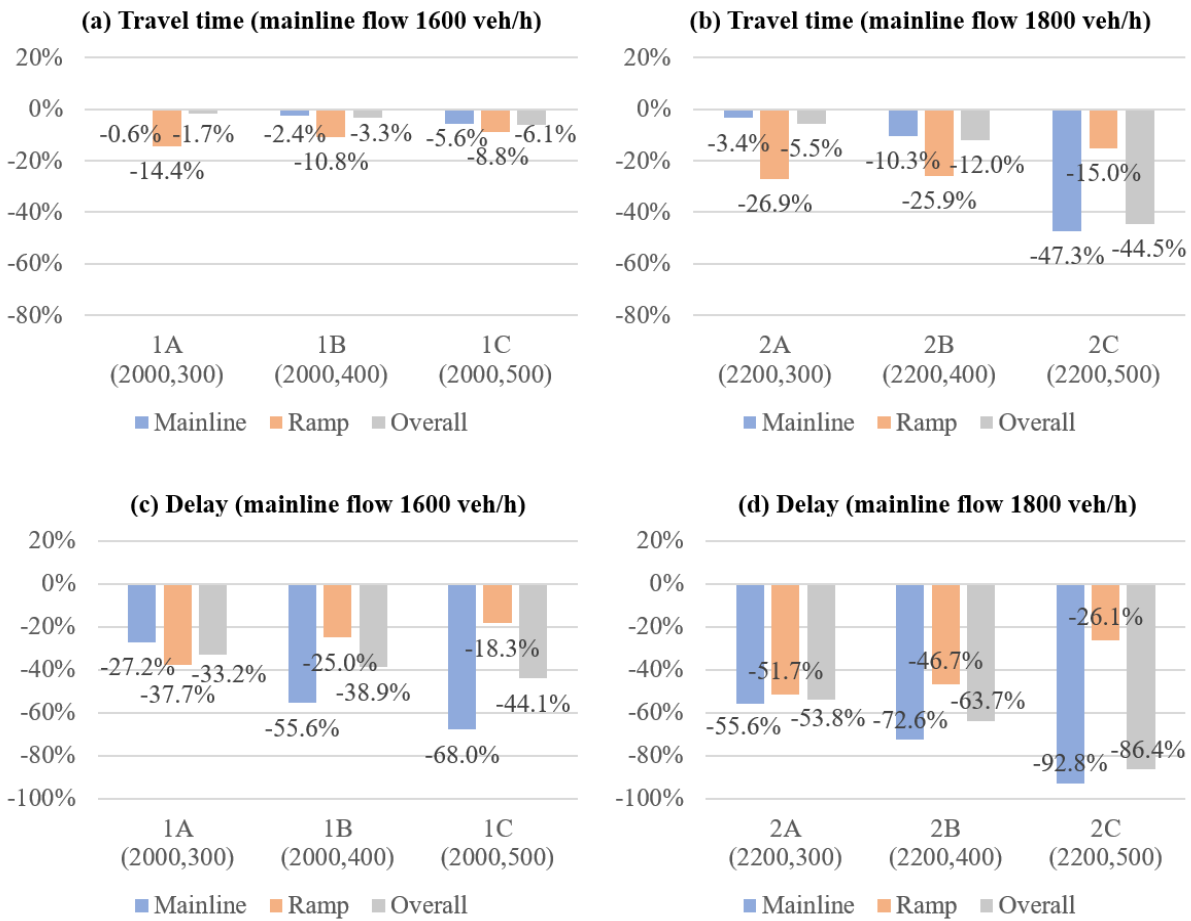

Fig. 5 Changes in travel time and delay

In Fig. 6, we illustrate how the aggregated vehicle speed changes across space and time in different scenarios. The speed data are collected at five-minutes intervals on the road segment between 500 meters and 2000 meters along the mainline freeway at a space resolution of 100 meters. This range covers the merging area (i.e., between 2000 to 2240 meters) and the upstream and downstream influence areas as defined in TRB (2016). The results are aggregated over all lanes (i.e., three lanes in the merging area and two lanes in the upstream and downstream segments) and all simulation runs of a scenario. According to the results, the CoMC case outperforms the base case at all demand levels in terms of less reductions in the vehicle speed. With CoMC, the speed remains above 95 km/h across the whole roadway network and the entire simulation period. In the base case of the most critical 2C scenario, the pattern of speed reduction persists and extends backwards along the main road (see Fig. 6k). The speed even drops below the critical mainline speed in the later stage of the simulation, indicating an onset of traffic congestions. Such congestions are successfully prevented when CoMC is applied (Fig. 6l). In addition, it is noted that the speed reduction in the base cases mainly originates from the merging area and spreads upstream and downstream, whereas in the CoMC cases, the speed reduction is more evenly distributed in the merging area and the upstream segment corresponding to the control range. This implies that CoMC can essentially shift part of the disturbances in the merging area to the nearby road segments, thereby reducing the intensity of the negative impacts.

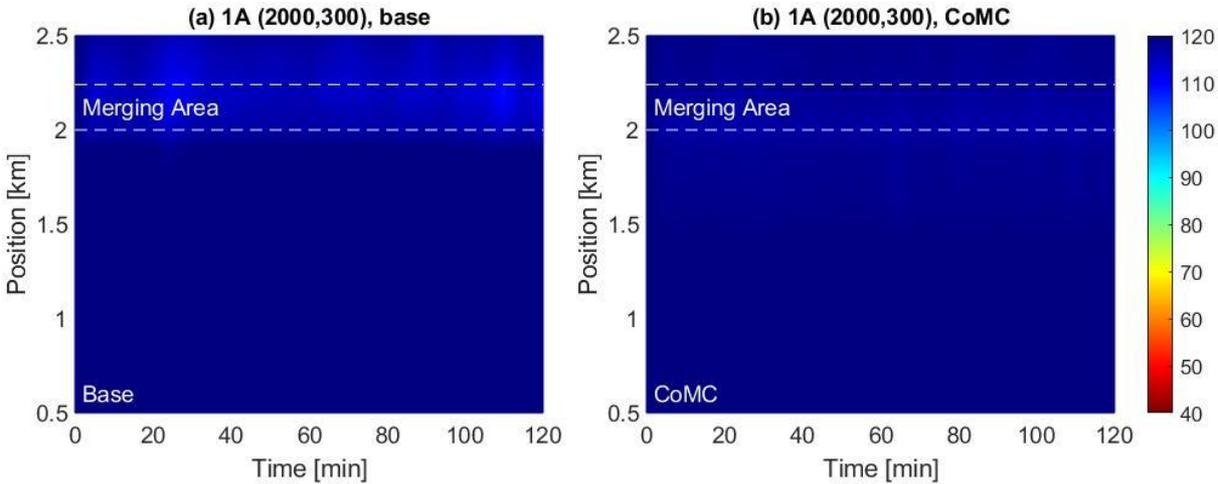

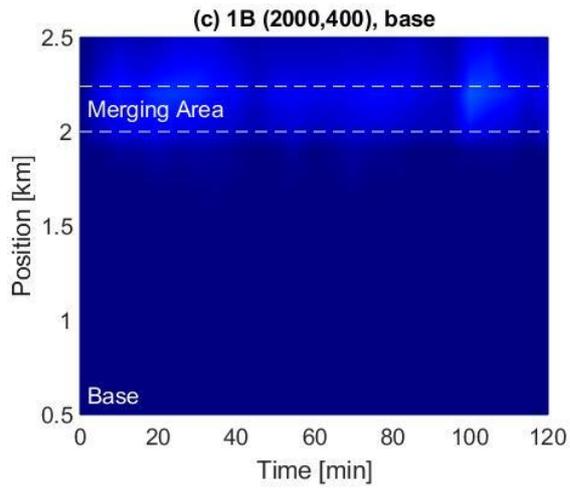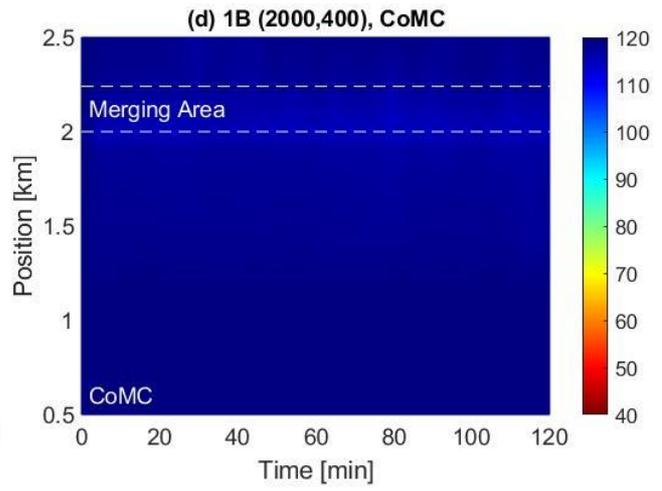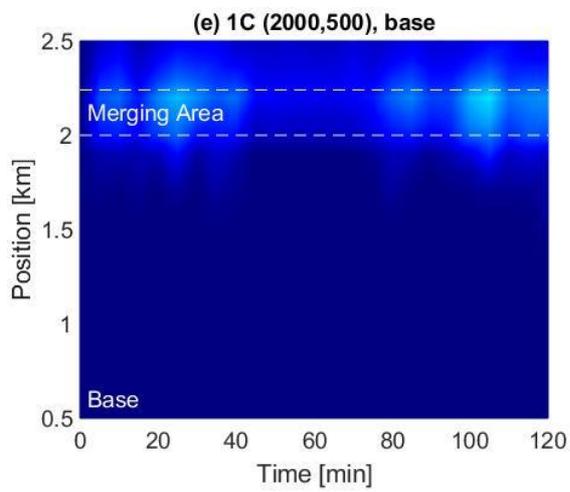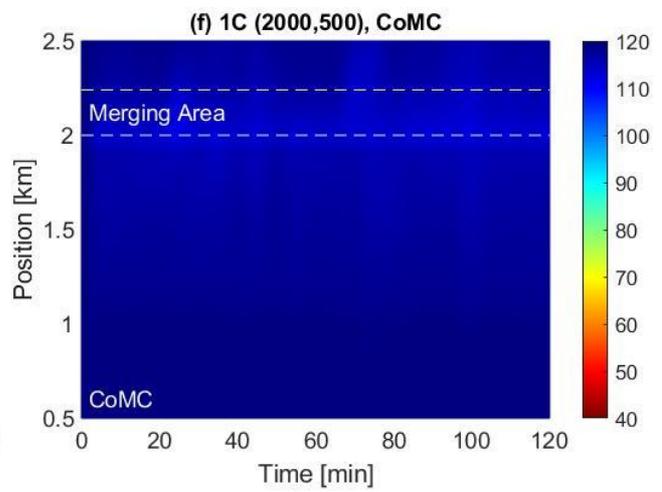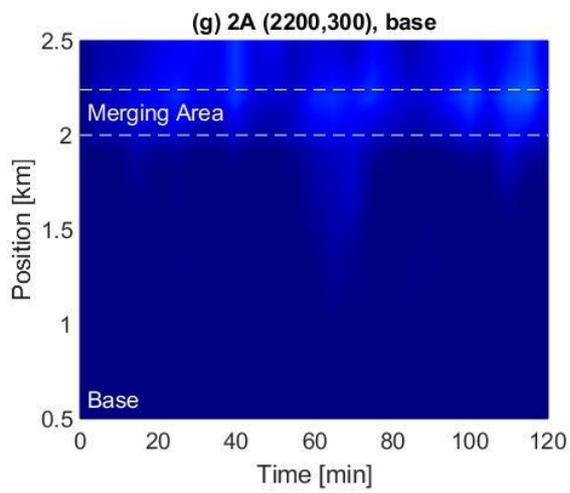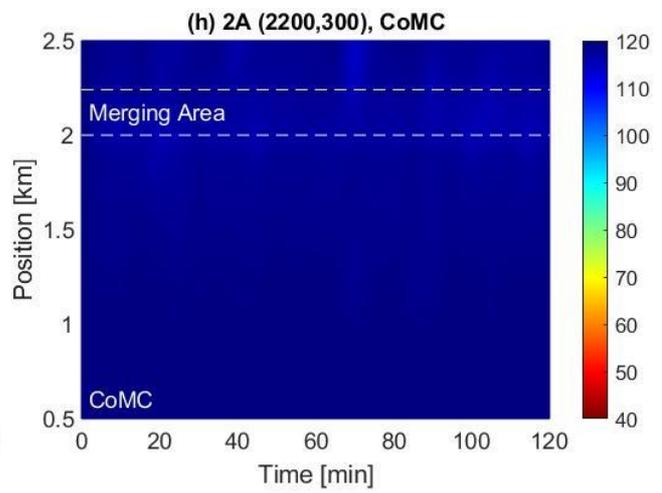

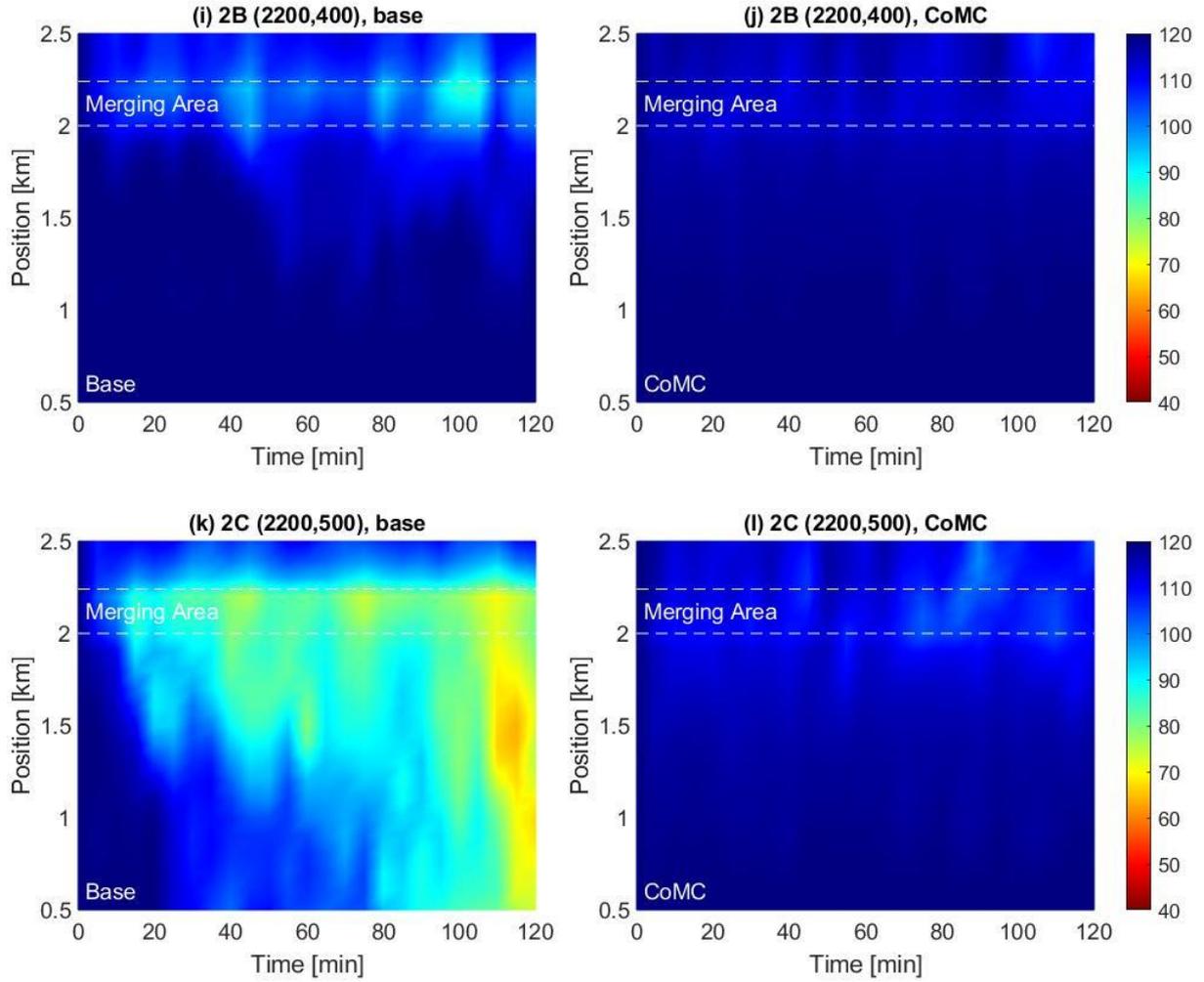

Fig. 6 Speed contour

## 5 Conclusion

In this paper, we present a flow-level CAV coordination strategy, which is a significant extension to Zhu et al. (2021b), to facilitate the merging operation in multilane freeway on-ramp bottlenecks. The strategy considers lane-changes between mainstream lanes and coordinates the ramp merging traffic with the mainline traffic in the outermost lane through proactive gap creation and platoon merging. One-sided lane-change prohibition rule is integrated to protect the created outer lane gaps from being occupied by the inner lane vehicles. The strategy is formulated as an optimization problem which determines the most efficient control plan adaptive to the real-time traffic state. Extended discussions on mainline-ramp priority and the proportion of lane-changes on the main road are provided. The efficiency of the proposed coordination is further demonstrated in a microsimulation-based case study where traffic from a one-lane on-ramp merge into a two-lane freeway. The results show that the coordination functions as expected in the simulated multilane environment and can substantially improve the traffic flow efficiency at ramp merging, especially under high traffic volume conditions. The main contributions of this paper are summarized as below: (1) it coordinates the two streams of traffic (instead of individual vehicles) for the flow-level efficiency gains at ramp merging; (2) it applies to the multilane freeway configuration which is more prevailing in the real world.

It is worth noting that in a multilane configuration, the lane-changing behaviors between mainstream lanes are important to the performance of traffic operation, as discussed in Section 3.3. Therefore, it would be beneficial to determine the optimal proportion of lane-changing vehicles, or more sophisticated, the exact vehicles to change lanes, based on the traffic conditions. This leaves an open question for future research. Further, the current strategy requires a 100% penetration rate of CAVs to ensure that any vehicle in the network can take the role of facilitating vehicle or platoon leader. This limit can be relaxed in the future research by adding a mechanism to assign the vehicles' roles based on their capabilities, so that the strategy can be applied to the mixed CAV-HDV traffic. In addition, as this study focuses on the formulation and validation of the proposed multi-lane strategy, assumptions on CAV capabilities, such as instantaneous communications and precise motion controls, are adopted. These should be further investigated in the future for an implementation in the real-world.

**Acknowledgement**

The authors are grateful to the Area of Advance Transport at Chalmers University of Technology for funding this research.